\documentclass[a4paper,twoside]{article}
\newcommand{\affil}[1]{$^{\rm #1}$}
\usepackage[authoryear]{natbib}
\usepackage{multirow}
\usepackage{amsmath,amsfonts,amssymb}
\usepackage{color}
\bibpunct{(}{)}{;}{a}{}{,}
\usepackage{graphicx}
\date{} 
\title{\large\bf\flushleft The Search for Supernova-produced Radionuclides in Terrestrial
	Deep-sea Archives}
\author{\parbox{\textwidth}{\flushleft
\vspace{-0.5cm}
{\it J. Feige\affil{A,G}, A. Wallner\affil{A,B,C}, S.~R. Winkler\affil{A}, S. Merchel\affil{D}, L.~K. Fifield\affil{B}, 
	 G. Korschinek\affil{E}, G. Rugel\affil{D}, and D. Breitschwerdt\affil{F}}\\
\vspace{0.4cm}
{\small \affil{A}\,University of Vienna, Faculty of Physics -- Isotope Research, 
	VERA Laboratory, W\"ahringer~Stra\ss e~17, 1090 Vienna, Austria}\\
{\small \affil{B}\,Department of Nuclear Physics, The Australian National University, Canberra, ACT 0200,\\ Australia}\\
{\small \affil{C}\,ANSTO, Locked Bag 2001, Kirrawee, DC, NSW 2232, Australia}\\
{\small \affil{D}\,HZDR, Bautzner Landstra\ss e 400, 01328 Dresden, Germany}\\
{\small \affil{E}\,Physics Department, TU Munich, James-Franck-Str., 85748 Garching, Germany}\\
{\small \affil{F}\,Department of Astronomy and Astrophysics, TU Berlin, Germany, Hardenbergstrasse 36, 
	10623~Berlin, Germany}\\
{\small \affil{G}\,Email: jenny.feige@univie.ac.at}}}
\begin{document}
\twocolumn[
\begin{changemargin}{.8cm}{.5cm}
\begin{minipage}{.9\textwidth}
\vspace{-1cm}
\maketitle
%
%

\small{\bf Abstract:} An enhanced concentration of $^{60}$Fe was found in a deep ocean's crust in 2004 in a layer corresponding to an age of $\sim$2 Myr. The confirmation of this signal in terrestrial archives as supernova-induced and detection of other supernova-produced radionuclides is of great interest. We have identified two suitable marine sediment cores from the South Australian Basin and estimated the intensity of a possible signal of the supernova-produced radionuclides $^{26}$Al, $^{53}$Mn, $^{60}$Fe and the pure r-process element $^{244}$Pu in these cores. A finding of these radionuclides in a sediment core might allow to improve the time resolution of the signal and thus to link the signal to a supernova event in the solar vicinity $\sim$2 Myr ago. Furthermore, it gives an insight on nucleosynthesis scenarios in massive stars, the condensation into dust grains and transport mechanisms from the supernova shell into the solar system.


\medskip
\medskip
\end{minipage}
\end{changemargin}
]
\small

\section{Introduction}

Stars with masses larger than 8 M$_\odot$ end their lifes in a supernova (SN) explosion. The nuclides, which were synthesized in the late phases of the star and during the explosion, are ejected and entrained in the supernova shell, which expands rapidly into the surrounding interstellar medium.
If such an event happens in the solar vicinity, the remnant may leave certain traces in terrestrial archives. 
Ejecta of a SN contain freshly produced long-lived radionuclides; eventually there might be a
chance of finding such radionuclides in terrestrial archives.

Such a signal of possible SN origin has been detected by \citet{knie} in the hydrogenetic ferromanganese crust 237KD, which was recovered from the equatorial Pacific. The radionuclide $^{60}$Fe with $t_{1/2}$ = (2.62 $\pm$ 0.04) Myr \citep{rugel} was measured with accelerator mass spectrometry (AMS) utilizing the MP-tandem accelerator in Munich. The signature was detected again in the same crust sample in the layers at a depth of 6-8 mm in a second measurement in the same laboratory, using different chemical separation schemes for AMS target preparation \citep{fitoussi}.

\begin{figure}[ht]
\begin{center}
\includegraphics[width=\linewidth]{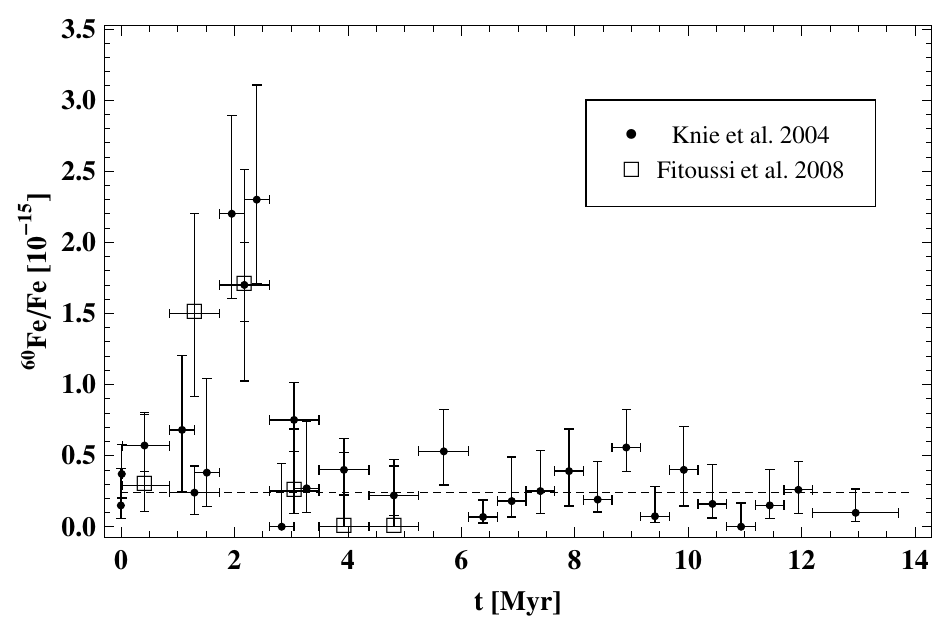}
\caption{Measured $^{60}$Fe/Fe ratios versus the age of the layer calculated with $t_{1/2}$($^{10}$Be) = 1.387 Myr. The black circles show the measurements by \citet{knie}, the white squares the data by \citet{fitoussi}.}\label{newhalflife}
\end{center}
\end{figure}

The age of the deep ocean's crust was determined with $^{10}$Be dating.
Based on the half-life of t$_{1/2}$($^{10}$Be) = (1.51 $\pm$ 0.06) Myr \citep{hofmann} \citet{knie} estimated a time range of 2.4-3.2 Myr for the layer, which contained the anomaly. \citet{fitoussi} measured the $^{10}$Be content consistently and reevaluated this result using $t_{1/2}$($^{10}$Be) = (1.36 $\pm$ 0.07) Myr \citep{nishiizumi}. The half-life of $^{10}$Be was recently remeasured by \citet{chmeleff} and \citet{korschinek} and the suggested value is now (1.387 $\pm$ 0.012) Myr. Following \citet{fitoussi} we calculate the time-profile with this new value and obtain a time range of 1.74-2.61 Myr. Figure \ref{newhalflife} shows the recalculated age profile of the sediment core including the data of \citet{knie} and \citet{fitoussi}.

A hint to recent SN activity in the solar vicinity is the existence of a thin hot cavity in the local interstellar medium (ISM), the Local Bubble (LB), which embeds our solar system. Its extensions are about 200~pc in the galactic plane and 600 pc perpendicular to it. Superbubbles are formed by several SN explosions, creating blast waves which push away the surrounding medium. They leave thin hot gas inside encircled by a shell mainly consisting of accumulated material, but also of SN ejecta. The latest age calculations of the LB, based on comparison between observations and simulations of ion column density ratios, suggest a value of 13.6-13.9~Myr \citep{avbr}.
\citet{fuchs} identified a stellar moving group to have crossed the solar neighborhood and determined about 14-20 SN explosions to form the LB, the last one occurring approximately 0.5 Myr ago. The remaining stars belong today to the subgroups UCL (Upper Centaurus Lupus) and LCC (Lower Centaurus Crux) of the Sco-Cen association. 

It is suggested that the stars, which created the LB, also contributed to the $^{60}$Fe excess on Earth \citep{bb02}. \citet{ralph} have searched for the birth place of neutron stars and their runaway companions to corroborate this theory.

To obtain further evidence for recent SN explosions in the solar vicinity,
we have identified two suitable marine sediments from the Indian Ocean, the Eltanin Piston Cores ELT 45-21 and ELT 49-53. Both sediment cores originate from a location southwest of Australia, 38$^\circ$58.7' S, 103$^\circ$43.1' E and 37$^\circ$51.6' S, 100$^\circ$01.7' E \citep{allled}. They were recovered from depths of 4237 and 4248 m in the years 1970 and 1971. With accumulation rates of 3-4 mm kyr$^{-1}$ the cores grow much faster than the ferromanganese crust ($\sim$2 mm Myr$^{-1}$). Such a high resolution allows a more precise dating of a SN signal compared to previous measurements and thus to constrain the time period of the incoming shock wave. In contrast to a crust the formation process of a sediment is well understood. The uptake-efficiency, the amount of an element to be incorporated into the sediment, might be up to 100~\%, in a manganese crust this value is only vaguely determined. 
If the SN signal is detectable in these two sediments, it allows to draw conclusions on the production rate of the measured radionuclides and transport mechanisms from the SN envelope to Earth.
The long-lived radionuclides $^{26}$Al, $^{53}$Mn, $^{60}$Fe, with half-lifes in the order of the SN-event $\sim$2 Myr ago, and the pure r-process element $^{244}$Pu are particularly applicable in this case.

\section{Summary of Previous Measurements}

More than 15 years ago it was suggested by \citet{ellis} and \citet{gunther96}, that there might be a chance for finding long-lived radionuclides in terrestrial archives, which were originally produced in a supernova -- a possible site for r-process. The first successful measurement was carried out by \citet{kniealt} in a ferromanganese crust from Mona Pihoa in the South Pacific. An enhanced $^{60}$Fe concentration was found in two layers corresponding to a total time range of 0-5.9 Myr, which was later confirmed with improved time resolution in the crust 237KD by \citet{knie} and \citet{fitoussi}. This signal was again observed in another crust (29DR-32) recovered from the North Pacific ocean (G. Korschinek, work in progress).

A measurement of $^{60}$Fe by \citet{fitoussi} of a marine sediment gave less promising results. The core taken from the North Atlantic ocean did not show a clear signal. \citet{fitoussi} suggested an overestimation of the fluence in the crust coming from the uncertainty of the uptake-factor. 
Samples from the Moon have been analyzed by \citet{cook} in terms of $^{26}$Al, $^{60}$Fe and $^{10}$Be. They state that one sample shows a signal, which can not be explained by the production of $^{60}$Fe by solar or galactic cosmic rays. However, the detected signal was one order of magnitude lower than expected.

The presence of another isotope, the r-process radionuclide $^{244}$Pu, was investigated at TU Munich in ferromanganese nodules \citep{wallner2000} and in a small piece of the same ferromanganese crust as used for the $^{60}$Fe measurements \citep{wallner2004}. Independently Paul et al.\ (2001, 2003, 2007) analyzed a marine sediment for possible r-process signatures. All these measurements did not result in a significant evidence of SN-produced $^{244}$Pu in such terrestrial archives. 
It is not clear whether only a fraction of SN explosions produce r-process $^{244}$Pu at all. The condensation and transport of $^{244}$Pu within dust particles in the SN shell into the solar system is also not understood sufficiently. 

Future AMS measurements of SN-produced $^{60}$Fe are planned by \citet{shawn}. By extracting $^{60}$Fe from magnetite microfossils, which are produced by magnetotactic bacteria, in deep-sea sediments, the search for a possible SN signal is extended to biogenic archives.

\section{Methods}

The red clay piston cores ELT 49-53 and ELT 45-21 might be suitable for detecting possible SN signatures. They were sampled at great depths in the Indian Ocean to minimize contamination from continental input. The age of the cores were already roughly determined with magnetostratigraphy and biostratigraphy \citep{allled}. Our sample of ELT 49-53 reaches from a depth of 120-517 cm, ELT 45-21 from 398-697 cm below the surface level of the ocean, covering the time range of 1.7-3.1 Myr, which include the time range of the $^{60}$Fe peak identified in the ferromanganese crust 237KD. Dating with $^{10}$Be gives further details on the age profile. The cores are split into sections of 1 cm length, corresponding to $\sim$3 kyr, with gaps of 8 cm in between, making it 71 samples in total.\\
Adequate chemical methods to extract Be, Al, Mn, Fe and Pu material (each element separately) from the sediment samples are described in \citet{bourles} and \citet{merchel}.\\
The most sensitive technique to date for detecting long-lived radionuclides is accelerator mass spectrometry (AMS). With this method the expected small amounts of such radionuclides in terrestrial archives can be quantified.

\section{Estimation of a Possible Signal in the Eltanin Cores}

\subsection{$^{10}$Be}

This isotope is mainly produced by spallation of oxygen and nitrogen by cosmic rays in the Earth's atmosphere. By assuming the cosmic ray flux was constant over the past few million years, it can be used to date the sections of the core with \citep{fitoussi}
\begin{equation}
 T_d = \frac{t_{1/2}(^{10}\mathrm{Be})}{\ln2}\ln\left(\frac{C_0}{C_d} \right),
\end{equation}
where $T_d$ is the age of the layer with depth $d$, $C_0$ the concentration of $^{10}$Be at the surface or any reference layer with known age, and $C_d$ the concentration at depth $d$.
An increase of $^{10}$Be by an enhanced cosmic ray flux from a SN would be small compared to the cosmogenic production. 
The yield of the ejected $^{10}$Be mass in a SN explosion is at least a factor of 10$^{3}$ lower than $^{60}$Fe \citep{ww95} and thus the SN input 2 Myr ago is estimated to be a factor of 10$^{3}$-10$^{4}$ lower than the cosmogenic fraction in a piece of sediment of 1 cm thickness. Therefore, quantification of any direct $^{10}$Be in the sediment cores above terrestrial cosmogenic $^{10}$Be is improbable.

\subsection{$^{26}$Al and $^{60}$Fe}

The radionuclide $^{26}$Al is dominantly produced by the $^{25}$Mg(p,$\gamma$)$^{26}$Al reaction in the core H burning phase of main sequence stars. A contribution in the later stages of massive stars occurs during the C and Ne shell burning and the explosive Ne burning. Its half-life is t$_{1/2}$~=~(0.717 $\pm$ 0.017) Myr, which is the weighted mean with standard deviation of the values reported in \citet{sam}, \citet{mid}, \citet{nor} and \citet{tho}, as stated in \citet{auer}. $^{60}$Fe is synthesized mainly by neutron capture on $^{59}$Fe in He and C shell burning and during the explosion, when the blast wave moves through the star, roughly in the same region where $^{26}$Al in produced \citep{limch}.

The $^{26}$Al and $^{60}$Fe signals in a deep-sea sediment core can be estimated by scaling from the $^{60}$Fe fluence estimation in the ferromanganese crust. The measured average ratio in the crust by \citet{knie} was $^{60}$Fe/Fe = 1.9 $\times$ $10^{-15}$ in the layers at a depth of 6-8~mm. This corresponds to a $^{60}$Fe fluence of $8.2 \times 10^5$ ats cm$^{-2}$ after background correction, which was an averaged $^{60}$Fe/Fe ratio of 2.4 $\times$ $10^{-16}$.
Two parameters remain uncertain. One is the uptake-factor $U$, as discussed above, the other parameter is the ratio of $^{60}$Fe/$^{26}$Al as it enters the solar system in the supernova remnant (SNR). \citet{knie} calculated an uptake-efficiency of 0.6 \%. However, a much higher value of 50-100 \% is suggested (G. Korschinek, work in progress). The $^{60}$Fe/$^{26}$Al flux ratio varies from 0.11-0.17 in observations from \textit{RHESSI} and SPI and from 0.16-1 in theoretical models \citep{limch}. With
\begin{equation}
 \left(\frac{^{60}\mathrm{Fe}}{^{26}\mathrm{Al}}\right)_m = \frac{^{60}\mathrm{Fe}}{^{26}\mathrm{Al}} U
	\times e^{-2.2 \ln2\left(\frac{1}{2.62}-\frac{1}{0.717}\right)} \label{hallo}
\end{equation}
the $^{26}$Al fluence is calculated, where $(^{60}\mathrm{Fe}/^{26}\mathrm{Al})_m$ is the ratio of the fluences one will measure today. Table~\ref{26Alfluence} shows the $^{26}$Al fluence values as expected in the sediments for different parameters of $U$ and the $^{60}$Fe/$^{26}$Al flux ratio, and table \ref{60Fefluence} the corresponding fluence values of $^{60}$Fe.

 \begin{table}[h] 
 \begin{center}
\caption{Estimated $^{26}$Al fluence $F_{26}$ (ats cm$^{-2}$) in ELT 45-21 and ELT 49-53 with varying uptake-factor $U$ and $^{60}$Fe/$^{26}$Al flux ratio.}\label{26Alfluence}
\small
\begin{tabular}{l|c|ccc}
\hline \multirow{2}{0.5cm}{$\frac{^{60}\mathrm{Fe}}{^{26}\mathrm{Al}}$} & \multicolumn{4}{c}{\hspace{0.9cm}$U$ [\%]} \\
& \multicolumn{2}{c}{\hspace{0.85cm}0.6} & 50 & 100 \\
\hline 0.11 & & 2.7 $\times$ 10$^8$ & 3.2 $\times$ 10$^6$ & 1.6 $\times$ 10$^6$ \\
0.5 & $F_{26}$  & 5.8 $\times$ 10$^7$ & 7.0 $\times$ 10$^5$ & 3.5 $\times$ 10$^5$ \\
1 & & 2.9 $\times$ 10$^7$ & 3.5 $\times$ 10$^5$ & 1.8 $\times$ 10$^5$ \\
\hline
\end{tabular}
\medskip\\
\end{center}
\end{table}

 \begin{table}[h]
 \begin{center}
\caption{Estimated $^{60}$Fe fluence $F_{60}$ (ats cm$^{-2}$) in ELT 45-21 and ELT 49-53 with varying uptake-factor $U$.}\label{60Fefluence}
\small
\begin{tabular}{lccc}
\hline $U$ [\%] & 0.6 & 50 & 100 \\
\hline $F_{60}$  & 1.4 $\times$ 10$^8$ & 1.6 $\times$ 10$^6$ & 8.2 $\times$ 10$^5$\\
\hline
\end{tabular}
\medskip\\
\end{center}
\end{table}

The number of atoms per cm$^2$ in the sediment cores is calculated with
\begin{equation}
 F_s = \frac{w N_A \rho h}{A}, \label{sflu}
\end{equation}
where $w$ is the mass fraction and $A$ the mass number of the isotope and $N_A$ Avogadro's constant. The density of the cores were estimated to be $\rho = 1.5$ g cm$^{-3}$ and the thickness of the layer is $h = 1$ cm. Such a layer  contains a time interval of $\sim$3 kyr.
We assume the sediments contain $\sim$10~\%~wt of stable Al and $\sim$5~\%~wt of stable Fe,
thus with equation \eqref{sflu} we obtain 3.3 $\times 10^{21}$ ats cm$^{-2}$ Al and 
8.1 $\times 10^{20}$ ats cm$^{-2}$ of Fe. Therefore, we expect the $^{60}$Fe/Fe ratio of the signal to be in the range of 10$^{-15}$-10$^{-13}$, the $^{26}$Al/Al ratio between 5~$\times$~10$^{-17}$ and 10$^{-13}$. The detection limit for measuring $^{26}$Al/Al and $^{60}$Fe/Fe with AMS is $\sim$10$^{-16}$, making the most pessimistic values hard to measure. By applying a dedicated chemical sample preparation technique, it is possible to reduce the fraction of stable Al and achieve an isotope ratio 3-10 times higher \citep{fitrai}. 

In contrast to $^{60}$Fe, $^{26}$Al has a much higher natural background. It is produced either by spallation from cosmic rays on Ar in the Earth's atmosphere or in-situ in the sediment. The atmospheric production can on the one hand be obtained by scaling the $^{10}$Be concentration in the ferromanganese crust. Using an initial atmospheric ratio of $^{26}$Al/$^{10}$Be = 1.89 $\times 10^{-3}$ \citep{auer} the background is estimated to be $\sim$5 $\times 10^3$ ats cm$^{-2}$ in a section of the core of 1 cm height.  Since also the uptake of $^{10}$Be is not well known, this can be considered as a lower limit. On the other hand we use the atmospheric production rate of $^{10}$Be computed by \citet{masbee} of 0.0209~ats~cm$^{-2}$~s$^{-1}$ and derive 5~$\times$~10$^5$~ats~cm$^{-2}$ as an upper limit.
A source of in-situ produced $^{26}$Al is the $^{23}$Na($\alpha,n$)$^{26}$Al reaction. The $\alpha$-particles are generated in the natural decay chains of $^{232}$Th, $^{235}$U and $^{238}$U.
If the Na, U and Th content in the cores is similar to that in the ferromanganese crust, then the production will be in the order of 10$^4$~ats~cm$^{-2}$, which is an order of magnitude lower than the estimated upper limit of the atmospheric production.
The contribution from micrometeorites can be neglected, as was pointed out by \citet{fitrai}.

\subsection{$^{53}$Mn}

The yield of $^{53}$Mn ejected by a SN explosion is similar or up to an order of magnitude higher than $^{60}$Fe (\citet{ww95} and \citet{rauscher}) in the mass range of 11 - 25 M$_\odot$ stars (if we neglect the 13 M$_\odot$ star from \citet{ww95}). If we take equation \eqref{hallo}, replace $^{26}$Al with $^{53}$Mn, and apply the half-life of $^{53}$Mn with $t_{1/2}$ = (3.7 $\pm$ 0.4) Myr \citep{honim}, we obtain the results shown in 
Table~\ref{53Mnfluence}.

 \begin{table}[h] 
 \begin{center}
\caption{Estimated $^{53}$Mn fluence $F_{53}$ (ats cm$^{-2}$) in ELT 45-21 and ELT 49-53 with varying uptake-factor $U$ and $^{60}$Fe/$^{53}$Mn flux ratio.}\label{53Mnfluence}
\small
\begin{tabular}{l|c|ccc}
\hline \multirow{2}{0.5cm}{$\frac{^{60}\mathrm{Fe}}{^{53}\mathrm{Mn}}$} & \multicolumn{4}{c}{\hspace{0.9cm}$U$ [\%]} \\
& \multicolumn{2}{c}{\hspace{0.85cm}0.6} & 50 & 100 \\
\hline 0.1 & \multirow{2}{0.5cm}{$F_{53}$} & 1.6 $\times$ 10$^9$ & 2.0 $\times$ 10$^7$ & 9.8 $\times$ 10$^6$ \\
1 & & 1.6 $\times$ 10$^8$ & 2.0 $\times$ 10$^6$ & 9.8 $\times$ 10$^5$ \\
\hline
\end{tabular}
\medskip\\
\end{center}
\end{table}

$^{53}$Mn is also produced in cosmic dust by solar cosmic rays. An extraterrestrial flux was calculated by 
\citet{auerdiss} with $\sim$200~ats~cm$^{-2}$~yr$^{-1}$ of $^{53}$Mn, which sums up to a fluence of 
6~$\times$~10$^5$~ats~cm$^{-2}$ in one cm of our sediment cores.

\subsection{$^{244}$Pu}

The long-lived radionuclide $^{244}$Pu (t$_{1/2}$ = (81.1 $\pm$ 0.3) Myr, calculated from the $\alpha$ decay half-life of t$_{1/2}$ = (81.2 $\pm$ 0.3) Myr \citep{agg} and spontaneous fission: t$_{1/2}$ = (66 $\pm$ 2) Gyr \citep{hoho} by \citet{lac}) is a product of pure r-process in massive stars during a SN explosion.
The yield in core collaps supernovae is estimated to be of the order of $M_{ej}$ = 10$^{-8}$ M$_\odot$ for $^{244}$Pu \citep{lingenfelter}. The number of atoms per cm$^2$ that reach the Earth is now calculated by spreading the ejected mass over a spherical shell with the radius $r$, which is the distance to Earth \citep{fields}
\begin{equation}
 F = \frac{1}{4}\frac{M_{ej}}{4 \pi A m_u r^2},
\end{equation}
with $A$ the atomic mass and $m_u$ the atomic mass unit. Choosing a very rough estimation for the distance of a SN explosion of $r$ = 100 pc, we expect $F_{244} \approx$~10$^4$~cm$^{-2}$. This value is small compared to the expected $^{60}$Fe fluence, but with the most sensitive AMS facilities for measuring $^{244}$Pu, it might be possible to detect a SN signal.\\
If $^{244}$Pu is found in the sediment cores, the question remains, if it is directly deposited from a single supernova. With its long half-life $^{244}$Pu should be abundant in the ISM by the contribution of multiple nucleosynthesis events. The expanding SN shell accumulates the surrounding medium, which is the low-density medium of the Local Bubble. The hot temperature of the LB of $\sim$10$^6$~K \citep{bb02}, prohibits the formation of molecules and dust grains. Therefore, it is still an open question, how $^{244}$Pu is collected from the ISM and if it is condensed into dust grains to be deposited on Earth.

\subsection{Deposition time}

The deposition time of SN ejecta on Earth is calculated using the thin shell approximation. If we assume that the entire mass is accumulated in the shell, then its thickness $d$ can be estimated with $d/r \sim 1/12$ for a strong shock into a monoatomic gas (e.g.\ \citet{cc}). With a distance $r$ of the SN explosion of the order of 100 pc, the shell has a thickness of $\sim$8 pc. The SNR expands freely into the ambient medium until the ejected mass is comparable to the swept-up mass of the ISM, subsequently the velocity can be calculated with the Sedov-Taylor solution
(e.g.\ \citet{cc})
\begin{equation}
 \dot{r} = \frac{2}{5}\left(\frac{2 E}{\rho}\right)^{1/2}r^{-3/2}.
\end{equation}
The energy of a SN explosion is typically $E = 10^{51}$~erg and $\rho$ is the mass density of the LB plasma. This results in a value of $5 \times 10^7$ cm s$^{-1}$ at 100 pc, which is still a Mach-3 shock.
Since the medium inside the LB has a low particle density of $n = 5 \times 10^{-3}$ cm$^{-3}$ and a temperature of $T = 10^6$ K \citep{bb02} the pressure is non-negligible. The solution only holds, if the kinetic energy of the remnant exceeds the thermal energy of the medium. With the pressure $P = 2nkT$ in an ionized gas, where $k$ is the Boltzmann-constant, and the adiabatic index $\gamma = 5/3$ (e.g.\ \citet{cc})
\begin{equation}
 \frac{P}{\rho(\gamma-1)} < \frac{1}{2} \dot{r}^2.
\end{equation}
Therefore, we obtain for the deposition time $t_d = d/\dot{r}$ $\approx 15$ kyr. 
The thin shell approximation refers to the ISM, which is compressed into a shell by the expanding shock wave. The SN ejecta is accumulated behind this shell, but because $M_{ej} \ll M_{sw}$, where $M_{ej}$ is the ejected mass and $M_{sw}$ the accumulated ISM mass, we expect the time of the SN ejecta to overrun the solar system to be much lower than 15 kyr.
It is not clear, how the dust grains are transported into the solar system and finally deposited on Earth. These mechanisms might lead to a broadening of a possible signal. \citet{fields} suggest a deposition time of the order of $~$10 kyr.

\section{Conclusion}

We have obtained two sediment cores from the Indian ocean and estimated a possible signal from the SN-produced radionuclides $^{26}$Al, $^{53}$Mn, $^{60}$Fe and $^{244}$Pu. In addition, the long-lived radionuclide $^{10}$Be is used to date the sediments. The calculations show that it is possible to measure $^{26}$Al, $^{53}$Mn, 
$^{60}$Fe and $^{244}$Pu with accelerator mass spectrometry, but the cosmogenic background of $^{26}$Al and $^{53}$Mn might be challenging to clearly distinguish it from a SN signal. \\
Because the sediment cores are split into sections of 1~cm length which corresponds to a time range of $\sim$3~kyr with gaps of 8 cm in between, a chance is given that the signal could be located in one of those gaps, corresponding to $\sim$24 kyr. The combination of two independent sediments lowers this probability. \\

We will continue the search for SN-produced radionuclides in terrestrial archives by analyzing the two deep-sea sediment cores with respect to the before mentioned SN-produced radionuclides. AMS measurements are planned at the VERA (Vienna Environmental Research Accelerator) facility (University of Vienna), ANU Canberra, Australia, TU Munich, Germany, HZDR Dresden, Germany, and ETH Zurich, Switzerland.\\
$^{26}$Al measurements with high time resolution will be performed and, with a half-life of 0.717 Myr, it is, like $^{60}$Fe, sensitive to single SN events. A positive signal will allow comparing its intensity with the observed $^{60}$Fe signal in the ferromanganese crust \citep{knie}. It provides a check of late stages in massive stars and in comparison with space-born observations of $\gamma$-ray intensities in our Galaxy (see e.g.\ \citet{roland}). \\
In parallel to $^{26}$Al, the SN-produced radionuclides $^{10}$Be, $^{53}$Mn and $^{60}$Fe will be measured in the same sections. $^{10}$Be is not sufficiently produced and ejected in a SN explosion, but it is constantly produced in the Earth's atmosphere and can be used to date the sections of the sediment cores.\\

The search for pure r-process $^{244}$Pu will be continued with new developments that improve the overall efficiency in AMS measurements, which was the prime limitation in the significance of the previous $^{244}$Pu data. 
While still only upper limits can be given for SN-produced $^{244}$Pu on Earth, the goal would be an unambiguous detection of extraterrestrial $^{244}$Pu, i.e., direct evidence that r-processes indeed happened within the last $\sim$100~Myr and can be identified.
The $^{244}$Pu signal was calculated for a SN occurring at a distance of 100 pc. It is very likely that the $^{60}$Fe signal identified by \citet{knie} and \citet{fitoussi} originated from a SN closer than 100 pc to the solar system \citep{ralph}, therefore we might be able to find a higher signal than estimated here. \\

If all of the above mentioned elements, which are all ejected in SN explosions, are detected, a clear link to their supernova-origin will be provided. The time resolution of the sediment cores is 1000 times higher than for the ferromanganese crust which will help to narrow the time range of the explosion in the solar vicinity. \\
It is not yet clear how dust is formed in a SN ejecta and how the grains are transported to Earth. The radionuclides are only able to overcome the ram pressure of the solar wind and the interplanetary magnetic field, if they are condensed into dust particles \citep{themis}. The dust condensation efficiency in SNRs is a critical value and not very well known yet. It ranges from 10$^{-4}$ in observations to 1 in theoretical simulations \citep{ouellette}. Comparing our measurements with theoretical SN models should give an insight on the formation of dust particles in SNRs and on nucleosynthesis scenarios in massive stars and during the explosion.

\section*{Acknowledgments} 
This work is funded by the Austrian Science Fund (FWF), project no. AI00428 and AP20434 through the European Science Foundation Collaborative Research Project CoDustMas.
We thank the Antarctic Research Facility, Florida State University, US (C. Sjunneskog) for generously providing the sediment cores and K. Knie (GSI Helmholtzzentrum f\"ur Schwerionenforschung GmbH, Darmstadt, Germany) for placing the $^{60}$Fe measurement data at our disposal.

\end{document}